\documentclass[online]{tlp}
\usepackage[T1]{fontenc}
\usepackage{graphicx}

\usepackage{xcolor}
\usepackage[nolist]{acronym}
\usepackage{pgfkeys}
\usepackage[listings,skins,breakable]{tcolorbox}
\usepackage{tikz}
\usetikzlibrary{arrows.meta, calc, positioning, decorations, matrix, trees, shapes}
\usepackage{siunitx}
\usepackage{xspace}
\usepackage{hyperref}
\usepackage{amsmath}
\usepackage{enumitem}

\definecolor{color1}{RGB}{239,111,106}
\definecolor{color2}{RGB}{99,136,180}
\definecolor{color3}{RGB}{85,173,137}
\definecolor{color4}{RGB}{140,194,202}
\definecolor{color5}{RGB}{187,118,147}

\pgfkeyssetvalue{/change colors/none}{red}
\pgfkeyssetvalue{/change colors/RB}{orange}
\pgfkeyssetvalue{/change colors/VMM}{green}

\newcommand{\changed}[2][none]{{\pgfkeysgetvalue{/change colors/#1}{\changecolor}\color{\changecolor}#2}}
\renewcommand{\changed}[2][none]{#2}

\acrodef{LLM}{Large Language Model}
\acrodef{SBFL}{Spectrum-based Fault Localization}
\acrodef{MBFL}{Mutation-based Fault Localization}
\acrodef{LLMFL}{Large Language Model Fault Localization}
\acrodef{SMT}{Satisfiability Modulo Theories}
\acrodef{AST}{Abstract Syntax Tree}
\acrodef{GRPO}{Group Relative Policy Optimization}
\acrodef{LoRA}{Low-Rank Adaptation}
\acrodef{SFT}{Supervised Fine-tuning}
\acrodef{ASP}{Answer Set Programming}

\newcommand{\profix}{\textsc{ProDebug}\xspace}
\newcommand{\gitseed}{\textsc{GitSEED}\xspace}
\newcommand{\gitlab}{\textsc{GitLab}\xspace}
\newcommand{\profl}{\textsc{ProFL}\xspace}

\NewTCBListing{prologlisting}{ !O{} }{%
  listing only,
  colback=black!10!white,
  colframe=black!15!white,
  hbox,
  leftrule=10pt,
  left=0pt,
  top=-2pt,
  bottom=-3pt,
  listing options={%
      numbers=left,
      basicstyle=\footnotesize\ttfamily,
      numberstyle=\color{black}\scriptsize,
      showstringspaces=false,
      breaklines=true,
      escapeinside={\%*}{*)},
      aboveskip=\smallskipamount,
      numbersep=6pt
    },
    #1} 

\newtcbox{\prologinline}[1][]{%
  on line,
  boxsep=2pt,
  boxrule=0.8pt,
  left=0pt,
  right=0pt,
  top=0pt,
  bottom=-0.5pt,
  colback=black!10!white,
  colframe=black!15!white,
  fontupper={\normalfont\ttfamily\footnotesize},#1}

\newcommand{\skchhole}{\begin{tikzpicture}[baseline={(n.base)}] \node[overlay,inner ysep=5pt,inner xsep=5pt,fill=white,rounded corners=2pt] (b) {}; \node[inner sep=1pt,outer sep=0pt] (n) at (b) {\normalfont ?};\end{tikzpicture}}

\tikzset{
  circ/.style={
    circle, draw=black, fill=white,
    inner sep=0pt, line width=0.4pt, font=\scriptsize,
    minimum size=2.0ex 
  }
}

\newcommand{\circledsm}[1]{\tikz[baseline=-0.8ex]{\node[inner sep=0.1em,circle,fill=color2,text=white] {\sffamily\small #1};}\xspace}

\newtcolorbox{summarybox}{
  colback=gray!5,
  colframe=gray!30,
  boxrule=0.3pt,
  arc=1pt,
  left=3pt,
  right=3pt,
  top=2pt,
  bottom=2pt,
  before skip=4pt,
  after skip=4pt
}

\usepackage{color}

\begin{document}

\lefttitle{Ricardo Brancas, Vasco Manquinho and Ruben Martins}

\jnlPage{1}{8}
\jnlDoiYr{2026}
\doival{10.1017/xxxxx}

\title[ProDebug: An Automated Debugging System for Prolog]{\textsc{ProDebug}: An Automated Debugging\\System for Prolog\thanks{This work was supported by Portuguese national funds through FCT, under projects UID/\-50021/\-2025 (DOI: {doi.org/\-10.54499/\-UID/\-50021/\-2025}), UID/\-PRR/\-50021/\-2025 (DOI: {doi.org/\-10.54499/\-UID/\-PRR/\-50021/\-2025}), 2023.14280.PEX (DOI: {doi.org/\-10.54499/\-2023.14280.PEX}) and through the Carnegie Mellon University Portugal Program under grant PRT/\-BD/\-152086/\-2021 (DOI: {doi.org/\-10.54499/\-PRT/\-BD/\-152086/\-2021}). This work was also partially supported by the National Science Foundation (NSF) under Award CCF2427581 and DARPA under Agreement FA8750-24-9-1000.}}

\begin{authgrp}
\author{\sn{Ricardo} \gn{Brancas}}
\affiliation{INESC-ID/Instituto Superior Técnico, Universidade de Lisboa, Portugal}
\author{\sn{Vasco} \gn{Manquinho}}
\affiliation{INESC-ID/Instituto Superior Técnico, Universidade de Lisboa, Portugal}
\author{\sn{Ruben} \gn{Martins}}
\affiliation{Carnegie Mellon University, USA}
\end{authgrp}

\history{\sub{xx xx xxxx;} \rev{xx xx xxxx;} \acc{xx xx xxxx}}

%
%
%
%
%
\maketitle              
\begin{abstract}
Prolog is a well-known declarative programming language commonly used in introductory courses on logic and reasoning. However, many students find Prolog challenging because it lacks the familiar debugging mechanisms found in imperative languages. In large classes, this difficulty is exacerbated by the challenge of providing timely and personalized feedback to students.

In this work, we introduce \textsc{ProDebug}, the first tool to combine Large Language Models (LLMs) with spectrum-based and mutation-based techniques for automated debugging of Prolog assignments. \textsc{ProDebug} automatically identifies faults and proposes bug repairs for student Git submissions. Faults are detected using three approach\-es---spectrum-based, mutation-based, and LLM reasoning---while repairs are generated using mutation-based techniques and LLMs.
Our evaluation on 1499 buggy student submissions from a bachelor's level programming class demonstrates the potential of automated, LLM-augmented feedback systems to scale support for declarative programming education.

\end{abstract}


\section{Introduction} \label{sec:introduction}

Debugging logic programs in Prolog poses a unique set of challenges. Traditional methods like tracing or inserting print statements are often insufficient, since Prolog's execution model is based on unification and backtracking rather than explicit control and data flows. This difficulty is particularly evident among learners, who often struggle to grasp the seemingly unpredictable flow of execution. While some debugging aids for Prolog have been proposed in the past~\citep{le2011incom,DBLP:conf/iclp/Neumerkel96a,DBLP:conf/issta/ThompsonS20}, they frequently rely on inflexible rule-based approaches and have cumbersome methods of interaction for students. As a result, many students rely on trial and error to detect and fix mistakes, making the learning process in Prolog and declarative programming more frustrating. To address this challenge, we introduce a tool that automatically analyzes Prolog programs, identifies possible bugs, and provides hints to help students correct their solutions.

Consider the task of defining a predicate \texttt{duplicate/2} that duplicates every element of a list. For instance, the query
\prologinline{?- duplicate([1,2],[1,1,2,2]).}
should succeed. Below is a student's submission for this problem.

\begin{prologlisting}
duplicate([], []).
duplicate([H|T], L2) :- duplicate(T, L1), L2 = (H,H,L1).
\end{prologlisting}

While the base case is correct, the recursive case is flawed. Instead of constructing a list, the student mistakenly creates a term of the form \prologinline{(H,H,L1)},
which does not represent the intended Prolog list structure. Our tool aims to automatically identify such errors and provide students with targeted hints. For this case, a possible hint, where \prologinline{\skchhole{}} highlights the part of the program the student should revise, could be:

\begin{prologlisting}
duplicate([], []).
duplicate([H|T], L2) :- duplicate(T, L1), L2 = %*\skchhole{}*).
\end{prologlisting}

\profix consists of two main components: a fault localizer, which identifies faulty clauses, and a repair module, which generates corrections for the identified bugs. A report is then generated and provided to the student based on this information.

The fault localizer supports three distinct methods: (1) \ac{SBFL}, (2) \ac{MBFL}, and (3) a method using \acp{LLM}. While concepts such as \ac{SBFL} and \ac{MBFL} are not new, our application to the Prolog context diverges significantly from previous approaches. Distinct from works that rely on mutation analysis to estimate coverage~\citep{DBLP:conf/issta/ThompsonS20}, \profix implements a novel \ac{SBFL} engine that leverages low-level Prolog tracing APIs. This allows us to capture exact execution spectra, including backtracking paths, which is critical for accurate localization.

Likewise, the repair module also supports multiple approaches:  a systematic mutation-based search using program synthesis~\citep{DBLP:conf/icst/BrancasMM25}, and a fine-tuned \ac{LLM}. To the best of our knowledge, our mutation engine is the first application of constraint-based, syntax-guided mutation to Prolog, enabling the enumeration of non-redundant mutations. Once a candidate repair is obtained, the differing parts of the programs are abstracted with question marks, producing hints that highlight where students need to revise their code.

This paper makes the following contributions:

\begin{itemize}
\item An automated fault localization approach for Prolog programs that integrates spectrum-based, mutation-based and \ac{LLM}-based approaches;
\item An automated repair procedure for Prolog that combines \ac{LLM}-based patching with synthesis-driven mutation;
\item The first fully automated tool for Prolog that offers both fault localization and program repair in an educational setting.\footnote{Available at \url{https://doi.org/10.5281/zenodo.18514417}.}
\end{itemize}

\section{System Architecture} \label{sec:system}

\begin{figure}[tb]
    \centering
    \scalebox{0.85}{\input{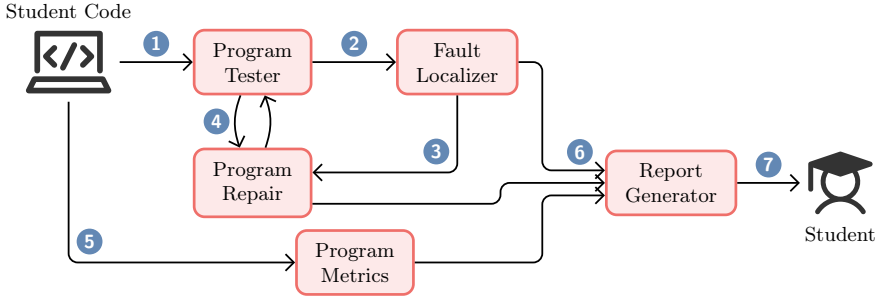}}
    \caption{System architecture overview.}
    \label{fig:system}
\end{figure}

\profix is an integrated tool designed to help Prolog students understand and fix problems in their coding assignments. \profix was developed as a Prolog extension for \gitseed~\citep{DBLP:conf/sigcse/OrvalhoJM24}, an open-source automated assessment platform.
When students submit their code to a \gitlab repository where \gitseed is activated, the student's program is automatically evaluated, and a feedback report is generated and returned to the student.

\autoref{fig:system} shows the overall system architecture of \profix. When a student's program is submitted, first it is tested against a suite of unit tests \circledsm{1}. If the program is incorrect (i.e., fails some test), it is passed to the fault localizer \circledsm{2} that attempts to find faults in the submission. Then, the fault report is sent to the program repair module \circledsm{3}, which uses that information to generate a patch to fix at least one of the program's bugs. This patch generation can take several attempts, and each attempt is tested against the test suite for correctness~\circledsm{4}. Complementary to the main fault localization and repair modules, \profix also collects additional program-level metrics and information \circledsm{5}. These include code complexity features such as the \emph{average clause length} or the \emph{number of clauses per predicate}.
The information from the fault localizer, repair module, and evaluation metrics is then collected into a report \circledsm{6}. This report can be customized by the faculty depending on the overall course design and the exercise's characteristics. Finally, the report is returned to the student \circledsm{7} through \gitseed.


\section{Fault Localization} \label{sec:fl}

One of the main features of \profix is the automatic identification of buggy Prolog code. We accomplish this through three different fault localization methods: \acf{SBFL}, \acf{MBFL} and \acf{LLMFL}. Next, we describe each method in detail.

\subsection{Spectrum-based Fault Localization}
\label{sec:sbfl}

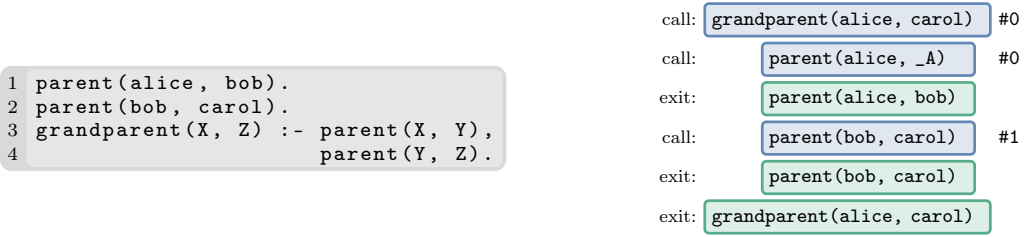
\begin{figure}
    \centering
\begin{minipage}{0.58\textwidth}
\begin{prologlisting}[right=0pt]
parent(alice, bob).
parent(bob, carol).
grandparent(X, Z) :- parent(X, Y),
                     parent(Y, Z).
\end{prologlisting}
\end{minipage}%
\begin{minipage}{0.50\textwidth}
\scalebox{0.85}{
\begin{tikzpicture}[very thick]
    \matrix[matrix of nodes, nodes={font=\ttfamily\small}, row sep=0.2em,
            column 1/.style={nodes={align=right,text width=4em,font=\normalfont\footnotesize}},
            column 2/.style={nodes={align=left,text width=13em}},
            column 3/.style={nodes={align=left,text width=6em}},
            call box/.style={draw=color2,fill=color2!20,rounded corners=0.2em},
            exit box/.style={draw=color3,fill=color3!20,rounded corners=0.2em},
            point box/.style={draw=color5,fill=color5!20,rounded corners=0.2em}]
    {
        call: & \node[call box] {grandparent(alice, carol)}; & \#0 \\
        call: & \node[call box,text width=9.5em,xshift=1em] {parent(alice, \_A)}; & \#0 \\
        exit: & \node[exit box,text width=9.5em,xshift=1em] {parent(alice, bob)}; & \\
        call: & \node[call box,text width=9.5em,xshift=1em] {parent(bob, carol)}; & \#1 \\
        exit: & \node[exit box,text width=9.5em,xshift=1em] {parent(bob, carol)}; & \\
        exit: & \node[exit box] {grandparent(alice, carol)}; & \\
    };
\end{tikzpicture}
}
\end{minipage}

\begin{minipage}{0.8\textwidth}

\end{minipage}%

    \caption{On the left of this figure we show a Prolog program, while on the right we show an execution trace for the query \prologinline{?- grandparent(alice, carol).}.}
    \label{fig:tracing}
\end{figure}

\acf{SBFL} is a widely known approach to finding faults in imperative programming languages \citep{DBLP:journals/tse/WongGLAW16}. Spectrum-based tools collect program execution traces (called \emph{program spectra}) from passing and failing test cases. By comparing which program elements (e.g., statements) are executed during successful runs versus failed runs, \ac{SBFL} assigns a suspiciousness score to each program element.

While spectrum-based methods are not directly applicable to declarative programming languages, due to the lack of an explicit control flow, we can exploit the mixed execution model of Prolog. Although Prolog programs have a declarative logical meaning, they are executed via a procedural mechanism based on unification, search, and backtracking, which can produce an observable trace.

Some Prolog interpreters, such as SWI-Prolog, have native support for extracting execution traces. These traces usually contain four types of information: \texttt{call} when a goal is started, \texttt{exit} when a goal is proved, \texttt{redo} when backtracking happens, and \texttt{fail} when a goal fails. 
In \autoref{fig:tracing}, we show an example of a Prolog program, the instrumented version of that program, and an execution trace for the query \prologinline{?- grandparent(alice, carol).}.
In \profix, a \emph{program element} corresponds to a Prolog clause (i.e., a rule or fact). Suspiciousness scores are thus assigned at the clause level, and spectrum information is extracted directly from the \texttt{call} and \texttt{redo} events of the trace.
%



Spectrum-based methods collect information from the execution of several tests. For each program element, \(s\), four pieces of information are collected: number of tests that passed and \(s\) was executed (\(e_p\)), number of tests that passed and \(s\) was not executed (\(n_p\)), number of tests that failed and \(s\) was executed (\(e_f\)), and number of tests that failed and \(s\) was not executed (\(n_f\)). The intuition behind the \ac{SBFL} method is that program elements that were disproportionately executed more often in failed test cases are more likely to be buggy. \changed[RB]{Crucially, the suspiciousness formulas compare \emph{relative} execution frequencies: a clause executed equally often across passing and failing tests receives a low score regardless of the absolute counts. In our setting, this is further supported by the use of test suites that include both positive tests (queries that should succeed) and negative tests (queries that should fail), ensuring that correct clauses are exercised in passing tests and providing a meaningful baseline for comparison.} Several formulas exist for turning these metrics into a suspiciousness score for each element. \profix implements some of the most commonly used ones~\citep{DBLP:conf/ijcai/ChatterjeeC0023}: Tarantula, Jaccard, Ochiai, Barinel, Kulczynski, Op2 and DStar.

\subsection{Mutation-based Fault Localization} \label{sec:mbfl}
\acf{MBFL} is a fault localization approach that leverages the principles of mutation testing~\citep{DBLP:conf/icst/MoonKKY14}. Mutation-based tools generate a set of \emph{mutants} by applying small syntactic changes to the program under test. These mutants are then executed against both passing and failing test cases to observe how the introduced changes affect program behavior. By analyzing which mutants are ``killed'' (detected by a test, meaning their behavior differs from the original program) and which survive, \ac{MBFL} computes a suspiciousness score for each clause. Clauses whose mutants are disproportionately killed by failing test cases are considered more likely to contain the fault~\citep{DBLP:conf/icst/MoonKKY14}.

\begin{figure}[tb]
\scalebox{0.87}{
    \begin{tabular}{ll}
        Original clauses:  \\[0.5ex]
        \prologinline{duplicate([], []).} & \prologinline{duplicate([H|T], L2) :- duplicate(T, L1), L2 = (H, H, L1).} \\[1ex]
        Example mutants:  \\[0.5ex]
        \prologinline{duplicate(\textcolor{color1}{\textbf{\_}}, []).} & \prologinline{duplicate([H | T], L2) :- duplicate(T, L1), L2 \textcolor{color1}{\textbf{is}} (H, H, L1).} \\
        \prologinline{duplicate([], \textcolor{color1}{\textbf{V0}}).} & \prologinline{duplicate([H | T], L2) :- duplicate(T, L1), L2 = (H, \textcolor{color1}{\textbf{\_}}, L1).} \\
        \prologinline{duplicate([], \textcolor{color1}{\textbf{1}}).} & \prologinline{duplicate([H | T], L2) :- duplicate(T, L1), L2 = \textcolor{color1}{\textbf{[}}H, H, L1\textcolor{color1}{\textbf{]}}.}
    \end{tabular}
    }
        \caption{Example of a program composed of two clauses and possible mutants generated by \profix for each of those clauses.}
    \label{fig:mutants}
\end{figure}


A central point of \ac{MBFL} is generating the program mutants. \profix accomplishes this through a logic-based mutation enumeration engine using \ac{SMT}~\citep{DBLP:series/faia/BarrettSST21}. This mutation enumeration engine encodes Prolog \acp{AST} as logic formulas and then relaxes parts of those formulas in order to enumerate mutations (a detailed technical description is provided in \ref{sec:appendix-smt}). 
While our approach is based on previous techniques for \ac{ASP} bug repair~\citep{DBLP:conf/icst/BrancasMM25}, adapting this method to Prolog requires addressing several key differences.
Unlike \ac{ASP}, where the order of rules and literals is irrelevant, Prolog's execution is sensitive to the order of clauses and body terms. Furthermore, the two languages differ significantly in operator and negation semantics (Negation as Failure in Prolog versus Stable Model semantics in \ac{ASP}). \profix also accounts for Prolog-specific features such as support for higher-order predicates, which are not present in \acl{ASP}.
\autoref{fig:mutants} shows some mutants generated by \profix for the two clauses of the introduction example program. Like for \ac{SBFL}, several suspiciousness formulas exist for \ac{MBFL}. \profix implements the two most commonly used ones: MUSE~\citep{DBLP:conf/icst/MoonKKY14} and Metallaxis~\citep{DBLP:journals/stvr/PapadakisT15}.

\subsection{Large Language Model-based Fault Localization} \label{sec:llmfl}

The third fault localization method included in \profix is \acf{LLMFL}. This approach takes advantage of \acp{LLM} trained on large amounts of data and especially code. Due to the specific deployment requirements of \profix (real-time usage in classroom with possibly few resources), we focus on fine-tuned open-access \emph{small} \acp{LLM} (\(\sim\! 4\) \changed[RB]{billion parameters}) in opposition to using closed-access models or very large instruction-tuned models.
Our prompt format, shown in \ref{sec:appendix}, includes a description of the program's desired functionality, a reference implementation provided by the faculty, and the faulty student submission. The output format consists of a sequence of buggy clauses identified in the student's code.

To fine-tune our fault localization models, we used \ac{LoRA}~\citep{DBLP:conf/iclr/HuSWALWWC22} and \ac{GRPO}~\citep{DBLP:journals/corr/abs-2402-03300}. \ac{LoRA} is a technique that uses rank decomposition matrices to greatly decrease the number of trainable parameters of a model. This technique has two main advantages: it reduces GPU VRAM requirements during training and it serves as a model regularizer, reducing the risk of overfitting. \ac{GRPO} is an online reinforcement learning method for training neural networks. In \ac{GRPO}, the model generates multiple candidate completions for each training example. A reward is then assigned to each completion using a user-defined reward function. Instead of optimizing each completion in isolation, \ac{GRPO} compares them against one another: the relative advantages between completions are used to guide updates to the model's policy, encouraging it to prefer higher-reward outputs. For more details on \ac{GRPO}, we refer to the original paper by~\cite{DBLP:journals/corr/abs-2402-03300}.

\section{Program Repair} \label{sec:repair}

The second main component of \profix is the program repair module. Given the fault localization report, its task is to find a fix for the bugs in the student submission. This fix can then be reported directly to the student or turned into a hint, depending on the faculty's preference. In this section, we describe two program repair approaches: one based on enumerating program mutations (\autoref{sec:repair-mutations}) and another leveraging \aclp{LLM} (\autoref{sec:repair-llms}).

\subsection{Mutation-based Repair}
\label{sec:repair-mutations}

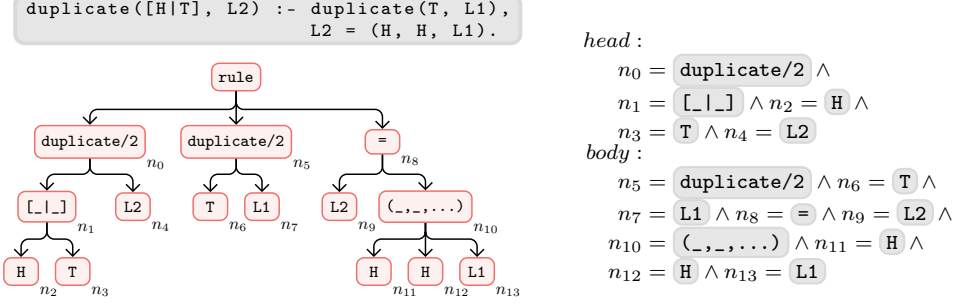
\begin{figure}[tb]
    \small
    \begin{minipage}{0.54\textwidth}
        \centering
        \begin{prologlisting}[listing options={%
      numbers=none,
      basicstyle=\scriptsize\ttfamily,
      showstringspaces=false,
      breaklines=true,
      escapeinside={\%*}{*)},
      aboveskip=\smallskipamount,
    },leftrule=1.5pt,right=0pt]
duplicate([H|T], L2) :- duplicate(T, L1),
                        L2 = (H, H, L1).
        \end{prologlisting}
        \scalebox{0.7}{\begin{tikzpicture}
    [thick,
        level distance=9ex,
        level 1/.style={sibling distance=8.5em},
        level 2/.style={sibling distance=5em},
        level 3/.style={sibling distance=3em},
        level/.style={arrows = {-Straight Barb[length=0.4em]},thick},
        tnode/.style={rounded corners=4pt,inner xsep=0.4em,inner ysep=0.5em,font=\ttfamily,minimum width=2em}, edge from parent/.style={draw,rounded corners=4pt},
        filled/.style={tnode,draw=color1,fill=color1!10},
        blank/.style={tnode,draw=black},edge from parent fork down]

    \node[filled] (root-1) {rule}
    child {node[filled] (n0) {duplicate/2}
            child {node[filled] (n1) {[\_|\_]}
                child {node[filled] (n2) {H}}
                child {node[filled] (n3) {T}}
            }
            child {node[filled] (n4) {L2}}
        }
    child {node[filled] (n5) {duplicate/2}
            child[sibling distance=3em] {node[filled] (n6) {T}}
            child[sibling distance=3em] {node[filled] (n7) {L1}}
        }
    child {node[filled] (n8) {=}
            child {node[filled] (n9) {L2}}
            child {node[filled] (n10) {(\_,\_,...)}
                    child {node[filled] (n11) {H}}
                    child {node[filled] (n12) {H}}
                    child {node[filled] (n13) {L1}}
                }
        };

    \begin{scope}[nodes = {below right = -3pt and -3pt}]
        \foreach \x in {0,...,13}{%
                \node at (n\x.south east) {$n_{\x}$};
            }
    \end{scope}

\end{tikzpicture}}
    \end{minipage}%
    \begin{minipage}{0.44\textwidth}
        \centering
        \footnotesize
        \addtolength{\jot}{-4pt}
        \begin{align*}
            head: \\
            n_0 & = \prologinline{duplicate/2}\,\land \\
            n_1 & = \prologinline{[\_|\_]} \land n_2 = \prologinline{H}\,\land \\
            n_3 & = \prologinline{T} \land n_4 = \prologinline{L2} \\
            body: \\
            n_5 & = \prologinline{duplicate/2} \land n_6 = \prologinline{T}\,\land \\
             n_7 & = \prologinline{L1} \land n_8 = \prologinline{=} \land n_9 = \prologinline{L2}\,\land \\
             n_{10} & = \prologinline{(\_,\_,...)} \land  n_{11} = \prologinline{H}\,\land\\
             n_{12} & = \prologinline{H} \land n_{13} = \prologinline{L1}
        \end{align*}
    \end{minipage}
    \caption{A buggy Prolog rule, its \acs*{AST} representation, and the \acs*{SMT} encoding for that tree.}
    \label{fig:mutation-ex}
\end{figure}

\profix's mutation-based repair is based on the \ac{MBFL} mutation engine introduced in \autoref{sec:mbfl}. \autoref{fig:mutation-ex} shows a concrete example of how \profix enumerates program mutations. First, the Prolog program (or program section) is parsed and transformed into an \acf{AST} (shown on the left side of \autoref{fig:mutation-ex}). Next, the \ac{AST} is used to create a logic formula that represents the program (shown on the right side of \autoref{fig:mutation-ex}). By relaxing some of the constraints in the formula and using an \ac{SMT} solver, we can generate new versions of the program with small differences (i.e., mutations). For instance, relaxing the constraint \(n_7 = \prologinline{L1}\) would allow the solver to enumerate the terms \prologinline{duplicate(T, L2)} or \prologinline{duplicate(T, H)}. \profix enumerates candidate repairs in increasing order of complexity: first one mutation, then two mutations, and so on.

Using an \ac{SMT}-based approach to generate mutations allows us to capture part of Prolog's semantics within the logical encoding. This enables \profix to constrain the search space by filtering out equivalent or semantically invalid mutations during enumeration. However, this expressiveness comes at the cost of increased computational overhead, as \ac{SMT}-based enumeration typically incurs higher runtime compared to purely syntactic mutation generation techniques.

Before creating the \ac{SMT} formula, the \ac{AST} can be artificially expanded in order to support more diverse mutations. For example, by adding two empty nodes as children of \(n_9\), we can enumerate the term \prologinline{[T | L1] = (H, H, L1)}, which would be impossible otherwise. \profix automatically \emph{completes} the AST by adding empty nodes so that all branches have the same depth and branching factor. Additionally, when the mutation engine is used for program repair, \profix enables some pruning constraints that help reduce the amount of equivalent and/or incorrect mutations. \ref{sec:appendix-smt} provides a complete description of the encoding, including the relaxation mechanism, \ac{AST} completion, and pruning constraints.

\subsection{Large Language Model-based Repair}
\label{sec:repair-llms}

Our \ac{LLM}-based repair approach uses a prompt that contains a short description of the student's assignment, a reference implementation created by the faculty, the student's incorrect submission and the list of clauses reported by the fault localization module (see \ref{sec:appendix}). As in the \ac{LLM}-based fault localization approach, we use small open-access \acp{LLM} to support classroom deployment.

For the repair fine-tuning, we used a combination of \ac{SFT} and \ac{GRPO}. As introduced in \autoref{sec:llmfl}, \ac{GRPO} creates several completions for each prompt and learns based on the relative rewards attributed to each of those completions. However, if the initial model performs the desired task very poorly, it can happen for all the completions to receive very similar rewards (and even all be 0 in extreme cases). In such cases, the model may learn very slowly or fail to learn altogether. To address this problem, we performed a short run of classical \acl{SFT} and then continue training with \ac{GRPO} from there. Like for the \ac{LLM} fault localization approach, we used \ac{LoRA} to decrease VRAM requirements during training and to serve as a regularizer.

\section{Evaluation} \label{sec:evaluation}

We focus the evaluation of \profix on a large set of correct and incorrect Prolog programs collected in a bachelor's level logic programming class. Overall, we considered 1499 instances, each consisting of a buggy and a corrected program.
Of these 1499 instances, 229 were collected from short practice exercises with an average of 2.9 clauses per exercise, while 1270 were collected from the longer final project assignment with an average of 33.3 clauses. \autoref{sec:methodology} explains in further detail how these instances were obtained.

In this evaluation, we aim to answer the following research questions:

\begin{enumerate}[leftmargin=*, label=\textbf{RQ\arabic*.}]
  \item How effective are the different fault localization methods?
  \item How do \profix's \ac{SBFL} and \ac{MBFL} compare with prior tools?
  \item How effective is the program repair?
  \item What is the impact of fine-tuning?
\end{enumerate}

\subsection{Methodology} \label{sec:methodology}

\subsubsection{Dataset}

The original set of programs used to create the instances in this evaluation was collected in a bachelor's level logic programming class~\citep{DBLP:journals/corr/abs-2504-16742}. Each data point in the dataset corresponds to a Git commit containing two versions of a student's program: the previous version and the current one. Among all commits, 2495 were identified as \emph{bugfix commits}, meaning the new version passed strictly more tests than the previous one.
Of these 2495, we filtered out commits with syntax errors and those where the ``fix'' only involved adding a missing predicate that had not been implemented in the previous version. Such cases were excluded because they do not represent true bug fixes. Students often build their programs incrementally, implementing new predicates over time. As such, a missing predicate is typically a feature not yet implemented, rather than an actual error. After this filtering, we obtained 1499 instances, each consisting of a buggy and a corrected program.

To evaluate the performance of the different fault localization methods, we need to establish the ground truth, i.e., which clauses in a program are faulty. This is challenging because students might modify clauses that are not relevant to the fix or might not fix all bugs in the program.
Therefore, we only include in the ground truth modified clauses that are relevant to the newly passed tests. We accomplish this by looking at the predicates invoked by these newly passed tests and computing their transitive closure over the program (i.e., all the predicates possibly called by the tests). The ground truth is then defined as the set of modified clauses that belong to these predicates.

\subsubsection{Fault Localization Evaluation}

The \ac{SBFL} and \ac{MBFL} approaches return a suspiciousness score for each clauses, where more suspicious clauses are assigned higher values. When several clauses are assigned the same suspiciousness, \profix uses the order they appear in the program to break ties, with elements that appear later being assigned higher ranks. The motivation for this is that students usually implement predicates sequentially and test as they go. As such, bugs are more likely to be present in later parts of the program.

\ac{LLM} fault localization methods directly return a ranking of suspicious lines. Lines not included in the ranking are placed below ranked lines in the inverse order they appear in the program, similarly to \ac{SBFL} and \ac{MBFL} disambiguation.

In the evaluation of the fault localization methods, we present three commonly used metrics~\citep{DBLP:journals/access/MaitamaIZ20}:

\begin{itemize}
    \item MinRank: first position in the ranking with a faulty clause;
    \item Accuracy@\(k\): percentage of faults in the program that appear in the first \(k\) positions of the ranking;
    \item Expense: percentage of the program a developer needs to go through (in the ranking order) until a fault is found. It is computed as the MinRank divided by the number of clauses.
\end{itemize}






\subsubsection{Program Repair Evaluation}

We consider a repair candidate correct if it flips at least one test to passing and does not flip any tests to failing. This follows the typical student use case where they only need a small hint in order to keep progress moving along. As such, when using mutation-based repair, the search terminates as soon as a correct repair candidate is found. However, inferencing several completions in the \ac{LLM} repair method only takes a little longer than a single one due to latency, batching and caching. As such, in this method, we generate several completions and give preference to ones that fix the most bugs in the program. If there are several tied candidate repairs in the number of tests flipped, we choose the one that is closest to the original student program. The current implementation of \profix's repair only supports repairing bugs at the clause-level and not term-level.

\subsubsection{\ac{LLM} Fine-tuning} \label{sec:impl-fine-tuning}

In order to fine-tune the \acp{LLM} for fault localization and repair, we need a large number of diverse instances with accurate ground truths. 
Therefore, we developed a bug insertion tool that can create realistic-looking bugs in correct Prolog programs. This tool uses a combination of rule-based modifications and the mutation engine introduced in \autoref{sec:mbfl} to insert different bugs in correct programs.
Based on the correct submissions from the dataset, we created 10,000 synthetic buggy instances to fine-tune the \acp{LLM}.

\subsection{Implementation}

\profix is implemented in Python and uses the SWI-Prolog dialect and interpreter through the Machine Query Interface. Parsing of Prolog programs is accomplished through a custom-built grammar. \profix uses a modified version of the Trinity framework \citep{DBLP:journals/pvldb/MartinsCCFD19} for the enumeration of mutations for fault localization, repair and creation of synthetic instances for fine-tuning. Results were obtained within a 10-minute time limit (wall clock time) and with 60GB of RAM per instance, which was strictly enforced through \texttt{runsolver}~\citep{runsolver} and \texttt{runhelper}.\footnote{\url{https://pypi.org/project/runhelper/}} We use a time limit of 10 minutes in this evaluation because we consider it the upper limit of time a student would accept to wait for automated feedback. Results were obtained on an Intel Xeon Silver 4210R processor. We employed the Qwen 3 family of models~\citep{DBLP:journals/corr/abs-2505-09388} for fault localization, and the Qwen 2.5 Coder family~\citep{DBLP:journals/corr/abs-2409-12186} for program repair. These model families were selected empirically based on their performance in preliminary experiments for each respective task. \acp{LLM} were fine-tuned on four Nvidia RTX A4000 GPUs, and inference was performed on a single RTX A4000. \profix's source code, data, and logs are publicly available.\footnote{\url{https://doi.org/10.5281/zenodo.18514417}}

\subsection{How effective are the different fault localization methods?} \label{sec:evaluation-fl}

\begin{table}[tb]
\caption{Different evaluation metrics for each fault localization method split among the practice exercises and the project submissions. \changed[RB]{Acc@\(k\): percentage of faults ranked within the top \(k\) positions (higher is better); Expense: MinRank divided by the total number of clauses, i.e., the fraction of the program inspected before the first fault is found (lower is better).}} \label{tab:fl-results-1}
\scalebox{0.85}{
\begin{tabular}{lrrrrrrr}
\topline
\textbf{Instance Type/Method} & \textbf{MinRank} & \textbf{Acc@1} & \textbf{Acc@3} & \textbf{Acc@5} & \textbf{Acc@10} & \textbf{Expense} & \textbf{Timeouts} \midline
\textbf{Practice Exercises} \small (n=229) \\
\hspace{1.5em}SBFL   & \textbf{1.23} & 83.57 & \textbf{97.73} & --    & --    & 48.29 & 4.4\% \\
\hspace{1.5em}MBFL   & 1.42 & 70.95 & 96.19 & --    & --    & 53.71 & 5.7\% \\
\hspace{1.5em}LLMFL & \textbf{1.23} & \textbf{84.75} & 97.01 & --    & --    & \textbf{47.53} & \textbf{0.0\%} \\
\hspace{1.5em}ProFL SBFL & 1.68 & 58.54 & 93.01 & --    & --    & 62.67 & 7.9\% \\
\hspace{1.5em}ProFL MBFL & 1.76 & 50.24 & 91.63 & --    & --    & 67.20 & 7.9\% \midline
\textbf{Project} \small (n=1270) \\
\hspace{1.5em}SBFL   & 3.33 & \textbf{56.82} & \textbf{63.22} & \textbf{71.35} & 82.74 & \textbf{10.86} & 9.1\% \\
\hspace{1.5em}MBFL   & 3.76 & 45.08 & 57.58 & 70.05 & \textbf{85.49} & 16.05 & 59.9\% \\
\hspace{1.5em}LLMFL & 6.32 & 24.11 & 41.06 & 55.77 & 74.85 & 19.11 & \textbf{0.0\%} \\ 
\hspace{1.5em}ProFL SBFL & \textbf{3.28} & 48.67 & 60.32 & 70.91 & 84.21 & 28.42 & 90.6\% \\
\hspace{1.5em}ProFL MBFL & 4.80 & 34.65 & 42.74 & 50.54 & 78.50 & 43.96 & 91.5\%
\botline
\end{tabular}
}
\end{table}

We evaluated the three fault localization methods proposed in Section~\autoref{sec:fl} and present the results in \autoref{tab:fl-results-1} under the labels \ac{SBFL}, \ac{MBFL} and \ac{LLMFL}. Of these methods, the one with the best overall performance when considering both types of instances is \ac{SBFL}, achieving 83.6\% Accuracy@1 in the Practice Exercises and 56.8\% Accuracy@1 in the Project submissions. This technique also has a fairly low percentage of timeouts\footnote{Timeouts in \ac{SBFL} result from copying large traces from SWI-Prolog to Python through a socket and do not represent a fundamental limitation of the approach.
} (\(8.4\%\)), meaning it can provide accurate answers in most instances. The second best performing technique is \ac{LLMFL}, with a very similar performance in the practice exercises. However, this approach struggles with the longer context of Project submissions, only achieving an Accuracy@1 of 24.1\%. The \ac{LLMFL} approach also has the big advantage of not having any timeouts. Finally, the worst performing approach is \ac{MBFL}, which is less accurate across the board. Furthermore, this approach has the drawback of having to generate and test large amounts of mutants, which scale linearly with the number of clauses in the program. This is reflected in the much larger percentage of timeouts in the \ac{MBFL} approach for the project (59.9\%) vs. for the practice exercises (5.7\%).

\begin{table}[tb]
    \caption{Fault localization performance across \ac{SBFL} and \ac{MBFL} formulas.}
    \label{tab:results-formulas}
    
    \setlength{\tabcolsep}{6pt}
    \begin{minipage}{0.5\textwidth}
        \centering
        \scalebox{0.85}{
        \begin{tabular}{lr}
        \topline
        \textbf{\ac{SBFL} Formula} & \textbf{Acc@1}  \midline
         Tarantula & 55.75\% \\
         \textbf{Ochiai} & \textbf{63.70\%} \\
         Op2 & 60.96\% \\
         Barinel & 55.75\% \\
         Jaccard & 63.57\% \\
         DStar & 50.31\% \\
         Kulczynski & 50.57\%
         \botline
    \end{tabular}
    }
    \end{minipage}%
    \begin{minipage}{0.5\textwidth}
        \centering
        \scalebox{0.85}{
        \begin{tabular}{lr}
        \topline
        \textbf{\ac{MBFL} Formula} & \textbf{Acc@1}  \midline
         \textbf{Metallaxis} & \textbf{52.87\%} \\
         MUSE & 50.29\%
         \botline
    \end{tabular}
    }
    \end{minipage}
\end{table}

\autoref{tab:results-formulas} shows summarized results for the \ac{SBFL} and \ac{MBFL} approaches using the different spectrum formulas implemented in \profix.
The best performing formula for \ac{SBFL} is Ochiai, with Jaccard as a close second. For \ac{MBFL}, the Metallaxis formula is slightly better than MUSE. Based on these results, \profix utilizes Ochiai and Metallaxis as its default formulas.

\begin{summarybox}
The most effective fault localization method is the \acl{SBFL}, followed by the LLM-based method and then the \acl{MBFL}. The \ac{SBFL} method correctly selects a buggy clause as the most suspicious in 84\% of practice exercises and 57\% of project submissions.
\end{summarybox}

\subsection{How do \profix's \ac{SBFL} and \ac{MBFL} compare with prior tools?}

\autoref{tab:fl-results-1} also includes the results of the \ac{SBFL} and \ac{MBFL} approaches implemented by a previous Prolog fault localization tool, \profl~\citep{DBLP:conf/issta/ThompsonS20}. Although both \profl and \profix implement \ac{SBFL} and \ac{MBFL} methods, their implementations differ substantially. We highlight three key distinctions: (1) \profix communicates with Prolog via the Prolog Machine Query Interface, rather than launching a new Prolog process for each query, (2) \profix supports a broader range of program mutations through its SMT-based mutation engine, and (3) \profix measures program coverage directly using Prolog’s tracing capabilities, whereas \profl estimates clause coverage indirectly through program mutations.
These differences address one of \profl's major limitations: its need to enumerate mutations even for the \ac{SBFL} approach, combined with launching a fresh process for each mutant, resulting in substantial execution overhead. Consequently, \profl exhibits a high rate of timeouts across both approaches---parti\-cu\-larly in the project submissions, where more than 90\% of instances exceed the 10-minute timeout used in our evaluation.

For the instances that finish within the time limit, \profix's \ac{SBFL} approach outperforms \profl's \ac{SBFL} on the practice exercises (83.6\% vs. 58.5\% Acc@1), while both tools show comparable performance on the project submissions (56.8\% vs. 48.7\% Acc@1, with \profl achieving a slightly better MinRank of 3.28 vs. 3.33). However, \profl's \ac{MBFL} approach performs consistently worse than \profix's \ac{MBFL} across all instance types and metrics, likely due to its more limited set of generated mutants. Notably, \profl's \ac{MBFL} achieves only 50.2\% Acc@1 on practice exercises compared to \profix's 70.9\%, and 34.7\% on project submissions compared to \profix's 45.1\%.

\begin{summarybox}
Overall, \profix represents a significant advancement in Prolog fault localization, improving both the accuracy of fault identification and the efficiency of the analysis compared to previous tools.
\end{summarybox}

\subsection{How effective is the program repair?}

\begin{table}[tb]
\caption{Evaluation results for \profix's program repair approaches.} \label{tab:repair-results}
\centering
\setlength{\tabcolsep}{6pt}
\scalebox{0.85}{
\begin{tabular}{lrr}
    \topline
    & \textbf{Practice Exercises} {\small (n=229)} & \textbf{Project} {\small (n=1270)} \midline
     \textbf{Simulated Perfect Fault Localization} \\
     \hspace{1.5em}Mutation & 21.0\% & 1.7\%  \\ 
     \hspace{1.5em}LLM w/ 100 completions (default) & 81.7\% & 34.3\% \\
     \hspace{1.5em}LLM w/ 30 completions & 79.0\% & 26.1\% \\
     \midline
     \textbf{\ac{SBFL} Fault Localization} \\
     \hspace{1.5em}Mutation & 20.2\% & 0.5\% \\ 
     \hspace{1.5em}LLM w/ 100 completions (default) & 73.8\% & 30.9\%  \\
     \hspace{1.5em}LLM w/ 30 completions & 69.0\% & 15.0\%
     \botline
\end{tabular}
}
\end{table}

To evaluate the effectiveness of our program repair methods, we tested them under two conditions: using a simulated \emph{perfect} fault localization and using the best-performing fault localization technique identified in \autoref{sec:evaluation-fl}. This design allows us to assess both the idealized repair performance, as well as to quantify the extent to which repair success is constrained by inaccuracies in fault localization. \autoref{tab:repair-results} summarizes the outcomes for the two repair strategies, mutation-based and \ac{LLM}-based, across these contexts.

Under simulated perfect fault localization, the mutation-based approach has a repair rate of 21.0\% on practice exercises and 1.7\% on project submissions. Although relatively low, these results are consistent with prior work on mutation-based repair for short declarative programs which achieved a 19\% repair rate~\citep{DBLP:conf/icst/BrancasMM25}. In contrast, the \ac{LLM}-based repair method performs substantially better, reaching 81.7\% for practice exercises and 34.3\% for project submissions. As described in \autoref{sec:impl-fine-tuning}, the LLM repair method samples multiple completions and selects the best candidate among them. Depending on deployment constraints, the number of completions can be reduced to improve runtime at a modest cost to repair accuracy. For example, reducing from 100 to 30 completions decreases average repair time from 106 seconds to 77 seconds, while lowering repair success by only 2.7 percentage points on exercises and 8.2 percentage points on projects.

When using the spectrum-based fault localization (\ac{SBFL}), repair performance decreases slightly. Mutation-based repair drops from 21.0\% to 20.2\% on exercises and from 1.7\% to 0.5\% on projects. The \ac{LLM}-based method also declines, from 81.7\% to 73.8\% on exercises and from 34.3\% to 30.9\% on projects. The reduction is more pronounced for the configuration using only 30 completions, especially on the project dataset, suggesting that diversity in completions helps compensate for fault localization errors.

\begin{summarybox}
Overall, when using the \ac{SBFL} localization, the \ac{LLM}-based repair achieves the best performance, repairing 73.8\% of practice exercises and 30.9\% of project submissions. These results compare favorably with the state of the art in declarative program repair and suggest that advances in fault localization could further improve repair success.
\end{summarybox}

\subsection{What is the impact of fine-tuning?}

\begin{table}[tb]
    \caption{Summarized results for fine-tuned and non fine-tuned \aclp{LLM}. The models used for the fault localization were Qwen 3 with sizes 4B, 8B and 14B, while the models used for repair were Qwen 2.5 Coder with sizes 3B, 7B and 14B.}
    \label{tab:finetune-effect}
    \centering
    \setlength{\tabcolsep}{6pt}
    \scalebox{0.85}{
    \begin{tabular}{lrr}
    \topline
         & \textbf{Fault Localization Acc@1} & \textbf{Repair Rate} \midline
       Fine-tuned LLM 4B/3B  & 33.4\%  & 34.2\%  \\ 
       Pre-trained LLM 4B/3B  & 14.2\% & 14.1\%      \\
       Pre-trained LLM 8B/7B  & 6.5\% & 21.5\%      \\
       Pre-trained LLM 14B  & 6.6\% & 37.9\%     
    \botline
    \end{tabular}
    }
\end{table}

In this work, we focused our \ac{LLM}-based approaches on relatively small models with around 4 billion parameters. To evaluate the limitations of these smaller models and assess the impact of fine-tuning, we compared our fine-tuned 4B/3B models with larger non–fine-tuned models from the same families. \autoref{tab:finetune-effect} summarizes these results.
As shown in \autoref{tab:finetune-effect}, fine-tuned models consistently outperform their non–fine-tuned counterparts of the same size in both fault localization and repair. Moreover, the fine-tuned 4B/3B models achieve results comparable to or even surpassing those of much larger (8B--14B) pre-trained models.

Interestingly, the non–fine-tuned Qwen 3 4B model performs substantially better than its 8B and 14B counterparts. According to the Qwen developers, the 4B version underwent a longer and more extensive training process than the other two variants, which likely explains its stronger baseline performance.

\begin{summarybox}
Fine-tuning has a significant positive impact on model performance. Our fine-tuned models (4B for fault localization and 3B for repair) achieve results that match or exceed those of much larger 14B pre-trained models within their respective families.
\end{summarybox}

\section{Related Work} \label{sec:related-work}

\paragraph{\textnormal{\textbf{Debugging Tools for Prolog}}}
%

Prolog's declarative semantics often confuse learners accustomed to imperative control flow. To address this, researchers have developed environments that prioritize reasoning about program logic over procedural execution. \textsc{GUPU} \citep{DBLP:conf/iclp/Neumerkel96a} provides a sandboxed environment using program slicing and test-based diagnostics to help students identify logical inconsistencies. Subsequent systems have focused on automated feedback: \textsc{PRAM} \citep{DBLP:conf/iticse/MansouriGH98} provides immediate correctness feedback via test suites, while \textsc{INCOM} \citep{le2011incom} uses weighted constraints to model student solutions and offer ranked hints based on common misconceptions.

Recent research has adapted imperative fault localization techniques to logic programming. \profl \citep{DBLP:conf/issta/ThompsonS20} applies spectrum-based and mutation-based analyses to rank clauses by suspiciousness, while Terminyzer \citep{DBLP:conf/padl/LiangK13} specifically targets non-termination by analyzing recursive call chains. While effective at localization, these frameworks lack automated repair capabilities or machine learning-driven reasoning.

\paragraph{\textnormal{\textbf{LLM-Augmented Debugging and Repair}}}
Recent research integrates \acp{LLM} with traditional debugging and repair methods. In declarative domains, \textsc{FormHe}~\citep{DBLP:conf/icst/BrancasMM25} combines logic-based fault localization, similarity metrics, program synthesis, and \ac{LLM} reasoning to locate and repair faults in \ac{ASP} programs. \changed[RB]{Both systems share the same \ac{SMT}-based mutation approach and overall design philosophy of pairing logic-based analysis with \ac{LLM} reasoning, though each system's mutation engine is adapted to its target language's syntax and semantics. However, they diverge in two key respects. For fault localization, \textsc{FormHe} exploits \ac{ASP}'s stable model semantics to identify minimal sets of rules responsible for a test failure, a technique with no counterpart in Prolog's procedural model; \profix instead introduces a novel \ac{SBFL} engine based on Prolog's native tracing APIs. For \ac{LLM} reasoning, \textsc{FormHe} fine-tunes models as classifiers, while \profix uses generative \acp{LLM} trained with \ac{GRPO}, and is evaluated at a substantially larger scale (1499 submissions versus 115). In a related direction,} \textsc{CodeHinter} \citep{DBLP:conf/kolicalling/Kurniawan0PNCJ25} demonstrates that \ac{LLM}-assisted feedback can improve debugging comprehension and learner confidence in introductory programming settings. These studies indicate that combining analytical debugging signals with \ac{LLM}-generated insights yields more interpretable and actionable feedback.

\profix extends this trajectory to Prolog, unifying spectrum-based, mutation-based, and \ac{LLM}-based reasoning to automatically localize and repair student faults. To our knowledge, \profix is the first tool to bridge traditional Prolog debugging with modern \ac{LLM} repair methods.

\section{Conclusion} \label{sec:conclusion}

This paper introduces \profix, an automated Prolog debugging system that combines spectrum-based, mutation-based, and \ac{LLM}-based reasoning for fault localization and repair. Our evaluation on 1499 student submissions demonstrates that \profix effectively addresses the challenge of providing automated debugging in declarative programming education.

Key results show that \acf{SBFL} achieved the highest localization accuracy, identifying buggy clauses in 80\% of practice exercises and 55\% of project submissions. For program repair, fine-tuned \acp{LLM} significantly outperformed mutation-based methods, reaching a 74\% success rate on practice exercises. These findings suggest that integrating logic-based analysis with generative reasoning provides a flexible debugging system suitable for classroom deployment. Future work will focus on multi-turn student interactions and leveraging term-level localization to further refine automated repairs.

\bibliographystyle{tlplike}
\bibliography{references}

@article{DBLP:journals/tse/WongGLAW16,
  author       = {W. Eric Wong and
                  Ruizhi Gao and
                  Yihao Li and
                  Rui Abreu and
                  Franz Wotawa},
  title        = {A Survey on Software Fault Localization},
  journal      = {{IEEE} Trans. Software Eng.},
  volume       = {42},
  number       = {8},
  pages        = {707--740},
  year         = {2016},
  doi          = {10.1109/TSE.2016.2521368},
  bibsource    = {dblp computer science bibliography, https://dblp.org}
}

@inproceedings{DBLP:conf/icst/MoonKKY14,
  author       = {Seokhyeon Moon and
                  Yunho Kim and
                  Moonzoo Kim and
                  Shin Yoo},
  title        = {Ask the Mutants: Mutating Faulty Programs for Fault Localization},
  booktitle    = {Seventh {IEEE} International Conference on Software Testing, Verification
                  and Validation, {ICST} 2014, March 31 2014-April 4, 2014, Cleveland,
                  Ohio, {USA}},
  pages        = {153--162},
  publisher    = {{IEEE} Computer Society},
  year         = {2014},
  doi          = {10.1109/ICST.2014.28},
  bibsource    = {dblp computer science bibliography, https://dblp.org}
}

@article{DBLP:journals/stvr/PapadakisT15,
  author       = {Mike Papadakis and
                  Yves Le Traon},
  title        = {{Metallaxis-FL}: mutation-based fault localization},
  journal      = {Softw. Test. Verification Reliab.},
  volume       = {25},
  number       = {5-7},
  pages        = {605--628},
  year         = {2015},
  doi          = {10.1002/STVR.1509},
  bibsource    = {dblp computer science bibliography, https://dblp.org}
}

@inproceedings{DBLP:conf/ijcai/ChatterjeeC0023,
  author       = {Prantik Chatterjee and
                  Jos{\'{e}} Campos and
                  Rui Abreu and
                  Subhajit Roy},
  title        = {Augmenting Automated Spectrum Based Fault Localization for Multiple
                  Faults},
  booktitle    = {Proceedings of the Thirty-Second International Joint Conference on
                  Artificial Intelligence, {IJCAI} 2023, 19th-25th August 2023, Macao,
                  SAR, China},
  pages        = {3140--3148},
  publisher    = {ijcai.org},
  year         = {2023},
  doi          = {10.24963/IJCAI.2023/350},
  bibsource    = {dblp computer science bibliography, https://dblp.org}
}

@article{DBLP:journals/corr/abs-2402-03300,
  author       = {Zhihong Shao and
                  Peiyi Wang and
                  Qihao Zhu and
                  Runxin Xu and
                  Junxiao Song and
                  Mingchuan Zhang and
                  Y. K. Li and
                  Y. Wu and
                  Daya Guo},
  title        = {DeepSeekMath: Pushing the Limits of Mathematical Reasoning in Open
                  Language Models},
  journal      = {CoRR},
  volume       = {abs/2402.03300},
  year         = {2024},
  doi          = {10.48550/ARXIV.2402.03300},
  eprinttype    = {arXiv},
  eprint       = {2402.03300},
  bibsource    = {dblp computer science bibliography, https://dblp.org}
}

@inproceedings{DBLP:conf/iclr/HuSWALWWC22,
  author       = {Edward J. Hu and
                  Yelong Shen and
                  Phillip Wallis and
                  Zeyuan Allen{-}Zhu and
                  Yuanzhi Li and
                  Shean Wang and
                  Lu Wang and
                  Weizhu Chen},
  title        = {LoRA: Low-Rank Adaptation of Large Language Models},
  booktitle    = {The Tenth International Conference on Learning Representations, {ICLR}
                  2022, Virtual Event, April 25-29, 2022},
  publisher    = {OpenReview.net},
  year         = {2022},
  url          = {https://openreview.net/forum?id=nZeVKeeFYf9},
  bibsource    = {dblp computer science bibliography, https://dblp.org}
}

@article{DBLP:journals/pvldb/MartinsCCFD19,
  author    = {Ruben Martins and
               Jia Chen and
               Yanju Chen and
               Yu Feng and
               Isil Dillig},
  title     = {Trinity: An Extensible Synthesis Framework for Data Science},
  journal   = {Proc. {VLDB} Endow.},
  volume    = {12},
  number    = {12},
  pages     = {1914--1917},
  year      = {2019},
  doi       = {10.14778/3352063.3352098},
  bibsource = {dblp computer science bibliography, https://dblp.org}
}

@article{runsolver,
  title = {Controlling a {{Solver Execution}} with the Runsolver {{Tool}}: {{System}} Description},
  shorttitle = {Controlling a {{Solver Execution}} with the Runsolver {{Tool}}},
  author = {Roussel, Olivier},
  year = {2011},
  month = nov,
  volume = {7},
  pages = {139--144},
  issn = {15740617},
  doi = {10.3233/SAT190083},
  journal = {Journal on Satisfiability, Boolean Modeling and Computation},
  number = {4}
}

@inproceedings{DBLP:conf/icst/BrancasMM25,
  author       = {Ricardo Brancas and
                  Vasco Manquinho and
                  Ruben Martins},
  title        = {Combining Logic and Large Language Models for Assisted Debugging and
                  Repair of {ASP} Programs},
  booktitle    = {{IEEE} Conference on Software Testing, Verification and Validation,
                  {ICST} 2025, Napoli, Italy, March 31 - April 4, 2025},
  pages        = {646--657},
  publisher    = {{IEEE}},
  year         = {2025},
  doi          = {10.1109/ICST62969.2025.10988950},
  bibsource    = {dblp computer science bibliography, https://dblp.org}
}

@inproceedings{DBLP:conf/iclp/Neumerkel96a,
  author       = {Ulrich Neumerkel},
  editor       = {Michael J. Maher},
  title        = {{GUPU:} {A} Prolog Course Environment and its Programming Methodology
                  (Poster Abstract)},
  booktitle    = {Logic Programming, Proceedings of the 1996 Joint International Conference
                  and Symposium on Logic Programming, Bonn, Germany, September 2-6,
                  1996},
  pages        = {549},
  publisher    = {{MIT} Press},
  year         = {1996},
  url          = {https://ieeexplore.ieee.org/xpl/articleDetails.jsp?arnumber=6278882},
  bibsource    = {dblp computer science bibliography, https://dblp.org}
}

@inproceedings{DBLP:conf/issta/ThompsonS20,
  author       = {George Thompson and
                  Allison K. Sullivan},
  editor       = {Sarfraz Khurshid and
                  Corina S. Pasareanu},
  title        = {{ProFL}: a fault localization framework for Prolog},
  booktitle    = {{ISSTA} '20: 29th {ACM} {SIGSOFT} International Symposium on Software
                  Testing and Analysis, Virtual Event, USA, July 18-22, 2020},
  pages        = {561--564},
  publisher    = {{ACM}},
  year         = {2020},
  doi          = {10.1145/3395363.3404367},
  bibsource    = {dblp computer science bibliography, https://dblp.org}
}

@inproceedings{DBLP:conf/padl/LiangK13,
  author       = {Senlin Liang and
                  Michael Kifer},
  editor       = {Konstantinos Sagonas},
  title        = {Terminyzer: An Automatic Non-termination Analyzer for Large Logic
                  Programs},
  booktitle    = {Practical Aspects of Declarative Languages - 15th International Symposium,
                  {PADL} 2013, Rome, Italy, January 21-22, 2013. Proceedings},
  series       = {Lecture Notes in Computer Science},
  volume       = {7752},
  pages        = {173--189},
  publisher    = {Springer},
  year         = {2013},
  doi          = {10.1007/978-3-642-45284-0\_12},
  bibsource    = {dblp computer science bibliography, https://dblp.org}
}

@inproceedings{DBLP:conf/iticse/MansouriGH98,
  author       = {Fatima Z. Mansouri and
                  Cleveland Augustine Gibbon and
                  Colin A. Higgins},
  editor       = {Gordon Davies and
                  M{\'{\i}}che{\'{a}}l {\'{O}}'h{\'{e}}igeartaigh},
  title        = {{PRAM:} prolog automatic marker},
  booktitle    = {Proceedings of the 6th Annual Conference on the Teaching of Computing
                  and the 3rd Annual {SIGCSE} Conference on Innovation and Technology
                  in Computer Science Education, ITiCSE 1998, Dublin City University,
                  Ireland, 18-21 August 1998},
  pages        = {166--170},
  publisher    = {{ACM}},
  year         = {1998},
  doi          = {10.1145/282991.283108},
  bibsource    = {dblp computer science bibliography, https://dblp.org}
}

@inproceedings{le2011incom,
  title={INCOM: A web-based homework coaching system for logic programming},
  author={Le, Nguyen-Thinh and Pinkwart, Niels},
  booktitle={Conference on Cognition and Exploratory Learning in Digital Age},
  pages={43--50},
  year={2011},
  organization={Citeseer}
}

@inproceedings{DBLP:conf/sigcse/OrvalhoJM24,
  author       = {Pedro Orvalho and
                  Mikol{\'{a}}s Janota and
                  Vasco Manquinho},
  editor       = {Mohsen Dorodchi and
                  Ming Zhang and
                  Stephen Cooper},
  title        = {{GitSEED}: {A} Git-backed Automated Assessment Tool for Software Engineering
                  and Programming Education},
  booktitle    = {Proceedings of the 2024 {ACM} Virtual Global Computing Education Conference
                  V. 1, {SIGCSE} Virtual 2024, Virtual Event, NC, USA, December 5-8,
                  2024},
  publisher    = {{ACM}},
  year         = {2024},
  doi          = {10.1145/3649165.3690106},
  bibsource    = {dblp computer science bibliography, https://dblp.org}
}

@article{DBLP:journals/access/MaitamaIZ20,
  author       = {Jaafar Zubairu Maitama and
                  Norisma Idris and
                  Abubakar Zakari},
  title        = {A Systematic Mapping Study of the Empirical Explicit Aspect Extractions
                  in Sentiment Analysis},
  journal      = {{IEEE} Access},
  volume       = {8},
  pages        = {113878--113899},
  year         = {2020},
  url          = {https://doi.org/10.1109/ACCESS.2020.3003625},
  doi          = {10.1109/ACCESS.2020.3003625},
  bibsource    = {dblp computer science bibliography, https://dblp.org}
}

@incollection{DBLP:series/faia/BarrettSST21,
  author       = {Clark W. Barrett and
                  Roberto Sebastiani and
                  Sanjit A. Seshia and
                  Cesare Tinelli},
  editor       = {Armin Biere and
                  Marijn Heule and
                  Hans van Maaren and
                  Toby Walsh},
  title        = {Satisfiability Modulo Theories},
  booktitle    = {Handbook of Satisfiability - Second Edition},
  series       = {Frontiers in Artificial Intelligence and Applications},
  volume       = {336},
  pages        = {1267--1329},
  publisher    = {{IOS} Press},
  year         = {2021},
  url          = {https://doi.org/10.3233/FAIA201017},
  doi          = {10.3233/FAIA201017},
  bibsource    = {dblp computer science bibliography, https://dblp.org}
}

@inproceedings{DBLP:journals/corr/abs-2504-16742,
  author       = {Ricardo Brancas and
                  Pedro Orvalho and
                  Carolina Carreira and
                  Vasco Manquinho and
                  Ruben Martins},
  title        = {Can Automated Feedback Turn Students into Happy Prologians?},
  booktitle    = {Proceedings 42nd International Conference on Logic Programming, {ICLP}
                  2026, Lisbon, Portugal},
  series       = {{EPTCS}},
  year         = {2026},
  url          = {https://doi.org/10.48550/arXiv.2504.16742},
  doi          = {10.48550/ARXIV.2504.16742},
  eprinttype    = {arXiv},
  eprint       = {2504.16742},
  bibsource    = {dblp computer science bibliography, https://dblp.org}
}

@article{DBLP:journals/corr/abs-2505-09388,
  author       = {Qwen Team},
  title        = {Qwen3 Technical Report},
  journal      = {CoRR},
  volume       = {abs/2505.09388},
  year         = {2025},
  url          = {https://doi.org/10.48550/arXiv.2505.09388},
  doi          = {10.48550/ARXIV.2505.09388},
  eprinttype   = {arXiv},
  eprint       = {2505.09388},
  bibsource    = {dblp computer science bibliography, https://dblp.org}
}

@article{DBLP:journals/corr/abs-2409-12186,
  author       = {Binyuan Hui and
                  Jian Yang and
                  Zeyu Cui and
                  Jiaxi Yang and
                  Dayiheng Liu and
                  Lei Zhang and
                  Tianyu Liu and
                  Jiajun Zhang and
                  Bowen Yu and
                  Kai Dang and
                  An Yang and
                  Rui Men and
                  Fei Huang and
                  Xingzhang Ren and
                  Xuancheng Ren and
                  Jingren Zhou and
                  Junyang Lin},
  title        = {Qwen2.5-Coder Technical Report},
  journal      = {CoRR},
  volume       = {abs/2409.12186},
  year         = {2024},
  url          = {https://doi.org/10.48550/arXiv.2409.12186},
  doi          = {10.48550/ARXIV.2409.12186},
  eprinttype   = {arXiv},
  eprint       = {2409.12186},
  bibsource    = {dblp computer science bibliography, https://dblp.org}
}

@inproceedings{DBLP:conf/kolicalling/Kurniawan0PNCJ25,
  author       = {Oka Kurniawan and
                  Erick Chandra and
                  Christopher M. Poskitt and
                  Yannic Noller and
                  Kenny Tsu Wei Choo and
                  Cyrille J{\'{e}}gourel},
  editor       = {Juho Leinonen and
                  Rodrigo Duran},
  title        = {Designing for Novice Debuggers: {A} Pilot Study on an AI-Assisted
                  Debugging Tool},
  booktitle    = {Proceedings of the 25th Koli Calling International Conference on Computing
                  Education Research, Koli Calling 2025, Koli, Finland, November 11-16,
                  2025},
  pages        = {41:1--41:7},
  publisher    = {{ACM}},
  year         = {2025},
  url          = {https://doi.org/10.1145/3769994.3769997},
  doi          = {10.1145/3769994.3769997},
  bibsource    = {dblp computer science bibliography, https://dblp.org}
}

\newpage

\appendix

\section{Large Language Model Prompts} \label{sec:appendix}

In this appendix we present the prompt templates used for the \acl*{LLM}-based fault localization and program repair approaches. Each prompt is split into the system prompt and the ``user'' prompt.

\begin{tcolorbox}[title=Fault Localization Prompt,fontupper=\ttfamily,fontlower=\ttfamily,breakable,enhanced]
Your task is to find faulty clauses in a Prolog program.

You should return a list of clauses in the program that are buggy.

Your final answer should only include exact copies of the clauses in the program which you think contain bugs and nothing else.
\tcblower
Here is the program description for this program:

\tcbox[size=fbox,colframe=white,colback=white]{<Program Description>}

Here is the reference implementation for this program:

\tcbox[size=fbox,colframe=white,colback=white]{<Reference Implementation>}

And here is the program where you need to identify the faulty clauses:

\tcbox[size=fbox,colframe=white,colback=white]{<Student's Program>}

\end{tcolorbox}

\begin{tcolorbox}[title=Program Repair Prompt,fontupper=\ttfamily,fontlower=\ttfamily,breakable,enhanced]
Your task is to repair faulty clauses in a Prolog program.

You should return a minimal diff of the program, such that the buggy program is correct after applying the diff.

You will be given a description of the problem, a correct reference implementation, a buggy implementation and a list of faulty clauses that should be corrected.

Your final answer should include the diff and nothing else.
\tcblower
Here is the program description for this program:

\tcbox[size=fbox,colframe=white,colback=white]{<Program Description>}

Here is the reference implementation for this program:

\tcbox[size=fbox,colframe=white,colback=white]{<Reference Implementation>}

And here is the program you need to correct:

\tcbox[size=fbox,colframe=white,colback=white]{<Student's Program>}

And these are the faulty clauses in the program:

\tcbox[size=fbox,colframe=white,colback=white]{<Faulty Clauses>}

\end{tcolorbox}

\section{SMT-based Mutation Engine} \label{sec:appendix-smt}

This appendix provides a detailed technical description of \profix's \ac{SMT}-based mutation engine, which underlies both the \ac{MBFL} fault localization (\autoref{sec:mbfl}) and the mutation-based repair (\autoref{sec:repair-mutations}). The engine is built on a modified version of the Trinity program synthesis framework~\citep{DBLP:journals/pvldb/MartinsCCFD19} and follows the same general approach used in our prior work on \ac{ASP} debugging~\citep{DBLP:conf/icst/BrancasMM25}. Trinity's original design enumerates programs from scratch given a domain-specific language (DSL) and input/output examples. In \profix, we repurpose the framework for \emph{mutation enumeration}: given an existing Prolog program, the engine systematically generates syntactic variants (mutants) by relaxing parts of the program's \ac{SMT} encoding. The main steps are: (1) defining a DSL that captures the space of valid Prolog terms, (2) parsing the student's program into an \ac{AST} and encoding it as an \ac{SMT} formula, (3) introducing relaxation variables that allow controlled deviations from the original program, and (4) iteratively querying the \ac{SMT} solver to enumerate mutants.

\subsection{Domain-Specific Language}

The mutation engine operates over a DSL that defines the syntactic constructs available for building Prolog programs. This DSL is instantiated per problem instance: it includes all predicates, operators, variables, and constants that appear in the student's program, augmented with a small fixed set of common Prolog built-ins. \autoref{fig:dsl} shows the core grammar.

\begin{figure}[tb]
\centering
\small
\begin{align*}
\mathit{stmt} &\to \mathit{clause}(\mathit{term},\; \mathit{body}) \mid \mathit{directive}(\mathit{term}) \\[3pt]
\mathit{body} &\to \mathit{and}(\mathit{term}^*) && \text{(body conjunction: \texttt{G\textsubscript{1}, G\textsubscript{2}, \ldots})} \\[3pt]
\mathit{term} &\to \mathit{predicate}/n\;(\mathit{term}^n) && \text{(user-defined + built-in predicates)} \\
              &\mid \mathit{op}_2\;(\mathit{term}, \mathit{term}) && \text{(binary: \texttt{=}, \texttt{is}, \texttt{<}, \texttt{;}, etc.)} \\
              &\mid \mathit{op}_1\;(\mathit{term}) && \text{(unary: \texttt{\textbackslash+}, \texttt{-})} \\
              &\mid \mathit{list}(\mathit{term}^*) && \text{(list literal: \texttt{[t\textsubscript{1}, t\textsubscript{2}, \ldots]})} \\
              &\mid \mathit{list\_c}(\mathit{term},\; \mathit{term}^*) && \text{(list with tail: \texttt{[t\textsubscript{2}, \ldots\;|\;t\textsubscript{1}]})} \\
              &\mid \mathit{tuple}(\mathit{term}^+) && \text{(tuple: \texttt{(t\textsubscript{1}, t\textsubscript{2}, \ldots)})} \\
              &\mid \mathit{curly}(\mathit{term}^*) && \text{(curly braces: \texttt{\{t\textsubscript{1}, t\textsubscript{2}, \ldots\}})} \\
              &\mid \mathit{atom} \mid \mathit{variable} \mid \mathit{integer} && \text{(leaf terms)} \\
              &\mid \texttt{\_} \mid \texttt{!} && \text{(don't-care and cut)} \\
              &\mid \mathit{empty} && \text{(absence of a node)}
\end{align*}

\caption{Core DSL grammar for \profix's mutation engine. Variable-arity constructs (\(\mathit{and}\), \(\mathit{list}\), \(\mathit{list\_c}\), \(\mathit{tuple}\), \(\mathit{curly}\)) use \(\mathit{empty}\) children to pad unused positions. The set of predicates, operators, variables, and constants is derived from the student's program.}
\label{fig:dsl}
\end{figure}

Each construct in the DSL is represented as a \emph{production} in the Trinity framework. Non-leaf productions have children (e.g., a predicate \texttt{append/3} takes three child terms), while leaf productions (e.g., the variable \texttt{X} or the integer \texttt{0}) have none.
A special production, \texttt{empty}, represents the absence of a node and is used to handle variable-arity constructs and optional children.

The set of available predicates is determined by analyzing the student's submission. For each predicate \(p/n\) appearing in the program, the DSL includes a non-leaf production with \(n\) children. If the program uses \texttt{maplist}, the DSL additionally includes curried variants of all predicates (e.g., \texttt{append/2} alongside \texttt{append/3}), since \texttt{maplist} applies a predicate with implicit extra arguments.

\subsection{AST-to-SMT Encoding}

Given a clause to mutate, \profix first parses it into an \ac{AST} and then encodes each node as an integer \ac{SMT} variable. The encoding proceeds as follows.

\paragraph{Node variables.}
Each \ac{AST} node \(n_i\) is assigned an integer variable whose domain is the set of production identifiers \(\{0, 1, \ldots, |P|-1\}\), where \(P\) is the set of all productions in the DSL.
A constraint of the form \(n_i = p_j\) asserts that node \(i\) takes production \(j\). For example, (\(n_8 = \texttt{=}\)) in \autoref{fig:mutation-ex} indicates that node 8 is the equality operator.

\paragraph{Binding types.}
Each node is assigned one of three binding types:
\begin{itemize}
    \item \textbf{Bound}: the node is fixed to its original production (\(n_i = p_{\mathit{orig}}\)). This node cannot be mutated.
    \item \textbf{Semi-bound}: the node is initialized to its original production but may deviate. A boolean \emph{relaxation variable} \(r_i\) tracks whether the node was mutated: \(r_i \Leftrightarrow (n_i \neq p_{\mathit{orig}})\).
    \item \textbf{Unbound}: the node is free to take any type-compatible production.
\end{itemize}
When used for fault localization (\ac{MBFL}), nodes in the target clause are marked as \emph{semi-bound}, while nodes in other clauses remain \emph{bound}. When used for repair, the same distinction applies, but additional \emph{unbound} nodes may be introduced via \ac{AST} completion (described below).

\paragraph{Type constraints.}
Each function production specifies the expected types of its children. The encoding enforces that whenever a parent node takes a particular production, each child must take a production whose left-hand-side type matches the expected child type:
\[
    (n_{\mathit{parent}} = p) \;\Rightarrow\; \bigvee_{q \,\in\, \mathit{compatible}(p, k)} (n_{\mathit{child}_k} = q)
\]
where \(\mathit{compatible}(p, k)\) is the set of productions whose output type matches the \(k\)-th child type of production \(p\). This ensures that every enumerated mutant is a syntactically well-formed Prolog program according to the DSL grammar.

\subsection{AST Completion} \label{sec:ast-completion}

A key feature of the engine is \emph{\ac{AST} completion}: before encoding, the \ac{AST} is expanded by adding \texttt{empty} child nodes so that every node has the same number of children (equal to the maximum branching factor in the DSL) and every branch reaches the same depth.

This step is critical because it allows the solver to ``grow'' new sub-expressions in place of \texttt{empty} nodes. For instance, given a tuple term \prologinline{(H, H, L1)}, \ac{AST} completion adds empty children so that the solver can replace the tuple with a list \prologinline{[H, H | L1]}, a structurally different term that requires additional child nodes. Without completion, such mutations would be impossible because the original \ac{AST} lacks the necessary tree positions. The bracket-style substitutions visible in \autoref{fig:mutants} (e.g., replacing \texttt{(} with \texttt{[}) arise precisely from this mechanism: the completed tree makes the list-constructor syntax reachable as a candidate without any special-casing.

When used for program repair, \profix can also introduce additional body terms in a clause by adding extra semi-bound root nodes to the body of the statement. These additional body roots are each assigned an integer position variable that determines where in the body sequence the new term is inserted, allowing the solver to synthesize new goals while preserving the existing clause structure.

\subsection{Mutation Control via Relaxation Variables}

The number of simultaneous mutations is controlled through the boolean relaxation variables \(r_i\) associated with semi-bound nodes. To enumerate mutants with exactly \(k\) changes, \profix uses a pseudo-boolean constraint:
\[
    \sum_{i \,\in\, \mathit{semi\text{-}bound}} r_i = k
\]
This is encoded using a \texttt{PbEq} constraint in the \ac{SMT} solver, which is equivalent to a cardinality constraint. During \ac{MBFL}, \profix enumerates mutations with \(k=1\) (single mutations), while during repair, it iterates with increasing values of \(k\) (first one mutation, then two, and so on) to explore repairs in increasing order of complexity.

\subsection{Pruning Constraints}

To reduce the number of semantically invalid or redundant mutants, the encoding includes the following families of pruning constraints, among others:

\begin{itemize}
    \item \textbf{Predicate parentage}: user-defined predicates cannot appear as children of other user-defined predicates (e.g., \texttt{append/3} cannot be a direct argument of \texttt{member/2}). This prevents the generation of higher-order terms that would be ill-formed in standard Prolog.

    \item \textbf{Head restrictions}: the head of a clause must be non-empty and must use one of the predicates defined in the student's program. Similarly, body terms are restricted to use known predicates and top-level operators.

    \item \textbf{Contiguous body terms}: if a body term in position\(i\) is \texttt{empty}, all subsequent positions must also be \texttt{empty}. This prevents gaps in rule bodies.
\end{itemize}

\subsection{Enumeration and Blocking}

The engine enumerates mutants by repeatedly querying the \ac{SMT} solver. After each satisfying assignment (i.e., each mutant), a \emph{blocking clause} is added to prevent the solver from returning the same assignment:
\[
    \bigvee_{i} (n_i \neq v_i)
\]
where \(v_i\) is the value assigned to \(n_i\) in the current model. This is implemented efficiently using Z3's variable substitution mechanism with a precomputed blocking template, avoiding the overhead of constructing a new formula for each blocking clause. The solver then searches for a new satisfying assignment, producing the next mutant.

Each enumerated mutant is decoded back into a Prolog program by traversing the \ac{AST} and replacing each node variable with the production selected by the solver. The resulting program is then tested against the test suite to determine whether the mutation was ``killed'' (for \ac{MBFL}) or constitutes a valid repair (for program repair).

\end{document}